\documentclass[a4paper,fleqn]{cas-sc}
\usepackage[numbers,sort&compress]{natbib}
\usepackage[T1]{fontenc}
\usepackage{lmodern}
\usepackage[utf8]{inputenc}
\usepackage{amsmath,amssymb}
\usepackage{booktabs,tabularx,array}
\usepackage{graphicx}
\usepackage{microtype}
\hypersetup{hidelinks}
\hypersetup{pdfauthor={Konstantinos E. Kampourakis, Vyron Kampourakis, Vasileios Gkioulos, Sokratis Katsikas}}
\begin{document}
\shorttitle{DT-assisted ATT\&CK mapping and dependency reasoning}
\shortauthors{Kampourakis et al.}
\title[mode=title]{Digital Twin-Assisted Mapping of ICS Telemetry to ATT\&CK for ICS with Evidence-Driven Dependency Reasoning}

% ORCID: 0009-0000-8883-0735
\author[1]{Konstantinos E. Kampourakis}[orcid=0009-0000-8883-0735]
\cormark[1]
\ead{konstantinos.kampourakis@ntnu.no}
% ORCID: 0000-0003-4492-5104
\author[1]{Vyron Kampourakis}[orcid=0000-0003-4492-5104]
\ead{vyron.kampourakis@ntnu.no}
% ORCID: 0000-0001-7304-3835
\author[1]{Vasileios Gkioulos}[orcid=0000-0001-7304-3835]
\ead{vasileios.gkioulos@ntnu.no}
% ORCID: 0000-0003-2966-9683
\author[1]{Sokratis Katsikas}[orcid=0000-0003-2966-9683]
\ead{sokratis.katsikas@ntnu.no}
\affiliation[1]{organization={Norwegian University of Science and Technology},Department of Information Security and Communication Technology,city={Gj\o vik},country={Norway}}
\cortext[1]{Corresponding author.}

\begin{abstract}
Reconstructing adversarial behavior from Industrial Control System (ICS) telemetry is difficult because process observations reveal physical changes more directly than the actions that produced them. This paper presents a Digital Twin (DT)-assisted framework that extracts synchronized state changes, converts them into evidence-preserving descriptions, maps them to ATT\&CK for ICS through retrieval-augmented Large Language Model (LLM) reasoning, and constructs a typed dependency graph. Evaluation comprises a four-configuration ablation on nine held-out SWaT scenarios containing ten telemetry-evaluable ground-truth episodes, three independent generations of the principal mapping configurations, and external evaluation on BATADAL and WADI. Across the three SWaT generations, DT-enriched mapping produces fewer annotation-relative False Positive (FP) episodes in every run, with a mean per-run reduction of 34.1\%. Both configurations obtain a mean recall of 0.633, although recall varies between 0.50 and 0.70 and DT enrichment does not improve F1 in every run. These observations are descriptive: the primary-run paired comparisons do not reach statistical significance, and the mapping advantage does not transfer to either external dataset. An exploratory positive-only dependency benchmark recovers eight of nine documented Co-occurrence relationships with DT context. A separate end-to-end benchmark containing one positive pair and 25 negative controls exposes propagation of mapping errors into unsupported edges. The findings identify both the potential and limitations of DT context for semantic security interpretation, without establishing reliable autonomous attribution, general causal reconstruction, or practical analyst benefit.
\end{abstract}
\maketitle

\section{Introduction}
\label{S:intro}

Industrial Control Systems (ICS) and Operational Technology (OT) networks constitute the computational backbone of modern critical infrastructure. In these settings, a cyber compromise does not typically remain confined to the digital domain; it can propagate to the physical process and ultimately affect system availability, integrity, and safety~\cite{giraldo2018survey,rotibi2024winning}. This cyber-physical coupling fundamentally distinguishes ICS security from traditional Information and Communication Technology (ICT) security. It motivates the need for analysis methods that capture not only whether malicious activity occurred, but also how it unfolded across time and process, thereby enabling stakeholders to more effectively analyze, prioritize, triage, and respond to security incidents according to their operational impact and potential consequences. In this context, a central challenge is that existing monitoring and detection pipelines typically operate at the level of alerts, anomaly scores, or low-level process deviations~\cite{mr2021machine,koay2023machine,kampourakis2026between}. While such outputs may be useful for flagging suspicious behavior, they provide limited support for reconstructing the attacker's strategy, understanding the ordering of attack steps, or reasoning about which actions enabled subsequent compromise. In practice, multi-stage attacks against ICS are inherently relational: reconnaissance may precede command abuse, command abuse may alter process parameters, and parameter manipulation may culminate in process disruption or unsafe physical behavior. Therefore, we distinguish that what is needed is a representation that can lift raw telemetry into a structured account of adversarial behavior and the dependencies among its constituent steps.

In this vein, MITRE ATT\&CK for ICS offers a natural foundation for such a representation~\cite{attckics,al2024mitre}. Unlike generic ICT-centric taxonomies, ATT\&CK for ICS models adversarial tactics and techniques targeting OT and cyber-physical processes, including behaviors such as unauthorized command messages, parameter modification, and availability loss. As a standardized semantic framework, it can enable heterogeneous observations to be described in terms of adversarial behavior and not device-specific artifacts alone. However, automatically mapping raw ICS telemetry to ATT\&CK for ICS remains an arduous task. Specifically, ICS datasets are dominated by numerical time-series measurements and sensor, actuator, and controller states~\cite{kampourakis2026digital, kampourakis2026ics}. In contrast, automated ATT\&CK mapping research has predominantly focused on textual Cyber-Threat Intelligence (CTI), vulnerability descriptions, and other explicit descriptions of adversarial activity~\cite{li2024automated,liu2022threat}. In this direction, this paper proposes a Digital Twin (DT)-assisted framework for semantic interpretation of ICS telemetry and evidence-driven dependency reasoning. Namely, the DT maintains a synchronized representation of asset, process, sensor, and actuator states. Changes in the DT state are transformed into structured event descriptions and mapped to ATT\&CK for ICS techniques using a retrieval-augmented Large Language Model (LLM). The resulting time-indexed ATT\&CK hypotheses are then organized within an evolving typed dependency graph. Altogether, the proposed framework is intended to support analyst-oriented interpretation of adversarial hypotheses and investigation of how those hypotheses may be related as an incident develops.

In detail, the evaluation separates technique attribution from dependency reasoning. For this reason, the SWaT dataset~\cite{mathur2016} is used as the primary benchmark because its documented scenarios support a held-out mapping ablation and a conservative dependency evaluation. BATADAL~\cite{Taormina2018} and WADI~\cite{ahmed2017} are subsequently used as external mapping-validation datasets under the frozen retrieval and mapping configuration. The resulting representation supports analyst-oriented interpretation of ATT\&CK hypotheses and investigation of how those hypotheses may be related as an incident develops. That said, the objective is not to establish complete attacker provenance or physical causality from historian telemetry alone, but to organize available cyber-physical evidence into a traceable structure for subsequent investigation. The main contributions of this paper can be summarized as follows:

\begin{itemize}
    \item \textbf{DT-based cyber-physical contextualization.} A synchronized state representation that combines ICS assets, process variables, actuator and controller states, nominal behavior, temporal history, and documented operational relationships to contextualize security-relevant process changes.

    \item \textbf{Evidence-preserving ATT\&CK mapping.} A pipeline that converts DT state changes into structured semantic evidence and maps them to ATT\&CK for ICS using frozen hybrid retrieval and structured LLM inference while retaining the observations supporting each attribution.

    \item \textbf{Controlled evaluation of DT context and generation variability.} A four-configuration SWaT ablation, three independent generations of the principal mapping configurations, frozen external evaluation on BATADAL and WADI, and stage-wise failure analysis characterize the observed mapping differences and their limitations. The results identify conditional reductions in unsupported hypotheses without establishing statistically significant superiority or consistent cross-dataset benefit.

    \item \textbf{Exploratory dependency reasoning and error-propagation analysis.} A typed evidence-driven dependency representation for relating time-indexed ATT\&CK hypotheses, together with experiments showing both the value of DT context for identifying operational coordination and its tendency to propagate incorrect upstream mappings into plausible but unsupported graph edges.
\end{itemize}

The remainder of the paper is organized as follows. Section~\ref{sec:background} reviews the necessary background and related work. Section~\ref{sec:arch} presents the system architecture. Section~\ref{sec:pipeline} details the pipeline design. Section~\ref{sec:eval} describes the experimental setup and evaluation results. Section~\ref{sec:dis} discusses limitations and implications, and Section~\ref{sec:con} concludes the paper.

% ═══════════════════════════════════════════════════════════════════════════════
\section{Background and Related Work}
\label{sec:background}

The proposed framework lies at the intersection of three research directions that have largely developed independently: semantic mapping of security evidence to ATT\&CK, DT-assisted cybersecurity for Cyber-Physical Systems (CPS), and attack-graph modeling of multi-stage adversarial behavior. This section reviews ATT\&CK mapping, DT-assisted cybersecurity, and cyber-physical attack graphs before positioning the proposed framework against representative work.

\subsection{ATT\&CK for ICS}
\label{sec:background_attck}

MITRE ATT\&CK for ICS is an empirically curated knowledge base that systematizes adversarial behavior in ICS and OT environments~\cite{attckics}. Unlike enterprise-oriented taxonomies, it captures actions whose effects may propagate to control functions and physical processes~\cite{alexander2020mitre}. Therefore, its tactics include process-oriented objectives such as Inhibit Response Function, Impair Process Control, and Impact~\cite{ekisa2024leveraging}. At the technique level, behaviors such as T0836~\textendash~Modify Parameter and the T1692~\textendash~Unauthorized Message family describe how adversaries may alter control parameters or manipulate communications with industrial devices. In particular, T1692.001~\textendash~Command Message captures unauthorized control commands, whereas T1692.002~\textendash~Reporting Message captures spoofed telemetry or process-state reports.

Overall, ATT\&CK for ICS provides a technology-independent semantic layer between low-level observations and higher-level adversarial behavior. Nevertheless, mapping ICS telemetry to this layer is challenging because process data typically expose the effects of an attack rather than the adversarial action itself~\cite{fluchs2023evaluation}. Precisely, historians and controller logs record measurements, outputs, set points, and device states, but, for example, a changed valve state does not by itself prove that an unauthorized command was issued, nor does an anomalous sensor trajectory directly expose a falsified reporting message. Accordingly, the proposed framework maps security-relevant DT observations to ATT\&CK techniques while preserving the evidence supporting each interpretation. Namely, technique observations remain linked to their assets, process stages, timestamps, confidence values, evidence levels, and underlying state changes. Preserving this provenance is particularly important when the mapped techniques are incorporated into dependency graphs, since erroneous semantic attribution could otherwise produce a coherent but unrealistic attack narrative.

\subsection{Semantic ATT\&CK Mapping}
\label{sec:background_mapping}

Automated ATT\&CK mapping has largely been approached as a natural-language understanding problem, using inputs such as CTI, vulnerability descriptions, and incident reports in which adversarial behavior is already expressed in security-oriented language. The objective is typically to associate textual evidence with one or more ATT\&CK tactics or techniques. For example, ATHRNN~\cite{liu2022threat} formulates ATT\&CK extraction from unstructured CTI as hierarchical multi-label classification, combining Transformer representations with an attention recurrent architecture that captures dependencies between tactics and techniques. Similarly, the work in~\cite{li2024automated} performs sentence-level tactic and technique extraction using DistilBERT-based multi-label classification together with post-processing based on ATT\&CK relationships. These approaches demonstrate the value of semantic representations and label dependencies, while also exposing challenges such as class imbalance, overlapping technique semantics, and limited context. However, both assume that the input already describes adversarial behavior in natural language.

ICS historian data pose an earlier semantic problem. Instead of stating that an adversary issued an unauthorized command, modified a parameter, or falsified a report, historians record measurements, set points, controller outputs, and equipment states from which such interpretations must be inferred. Therefore, the challenge lies not only in mapping evidence to ATT\&CK, but first in transforming numerical cyber-physical behavior into evidence suitable for semantic interpretation. In this regard, the proposed framework performs this translation through a DT. Specifically, synchronized cyber-physical state changes are converted into structured verbalizations that preserve the affected asset, process stage, variable role, temporal evolution, deviation from normal behavior, and available operational context. Then, retrieval constrains a general-purpose LLM to relevant ATT\&CK for ICS candidates. The resulting technique observations remain time-indexed and linked to their supporting evidence, enabling their subsequent use in dependency reasoning rather than reducing an incident to an unordered set of labels.

\subsection{DTs for ICS Cybersecurity}
\label{sec:background_dt}

DTs have increasingly been explored for ICS cybersecurity as they provide synchronized representations against which physical-system behavior can be assessed. Existing work spans secure state replication, virtualized attack experimentation, behavioral anomaly detection, and runtime verification. For instance, the work in~\cite{gehrmann2019digital} uses the DT as a protected state-replication environment for industrial automation, emphasizing synchronization, isolation, access control, and secure communication. This work establishes the importance of maintaining consistency between digital and physical states, but its focus is on secure replication and not semantic interpretation of observed changes. EPICTWIN~\cite{kandasamy2022electric} instead uses the DT as a scalable virtual environment for reproducing ICS devices, communication infrastructure, and attacks, supporting controlled and repeatable cybersecurity experimentation.

Other approaches exploit the DT as a behavioral reference. LATTICE~\cite{xu2023digital} combines DT modeling and curriculum learning for CPS anomaly detection and is evaluated on datasets including SWaT, WADI, and BATADAL. CRYSTAL~\cite{moradi2024crystal} integrates formal attack models, model checking, and a runtime Tiny DT to identify observations that violate permitted state transitions. Similarly, the study in~\cite{alcaraz2024digital} combines DT representations with machine-learning mechanisms to identify malicious deviations from expected industrial behavior, while PLC-MDT~\cite{xu2025plc} separately models sensor measurements and control signals to expose inconsistencies caused by control-logic modification and coordinated stealthy attacks. Across these approaches, the DT primarily determines whether observed behavior deviates from an expected or formally valid system state. Collectively, as we discern, prior work establishes DTs as useful foundations for secure synchronization, virtual experimentation, expected-state estimation, runtime verification, and anomaly detection. Less explored is their use as a contextual substrate for semantic incident reconstruction. In this respect, the framework proposed addresses this gap using synchronized DT state not only to identify deviations, but to determine which ATT\&CK for ICS behavior is supported by the available evidence and how multiple observations relate as an incident evolves.

\subsection{Attack Graphs for CPS}
\label{sec:background_graphs}

Attack graphs support reasoning about multi-stage compromise by representing vulnerabilities, privileges, connectivity, and attack preconditions and post-conditions. In CPS and ICS environments, this abstraction should additionally capture how cyber compromise interacts with physical-system behavior. Representatively, A2G2V~\cite{al2019a2g2v} automates attack-graph generation through model checking by combining network architecture, vulnerabilities, attack actions, state variables, and security properties to enumerate feasible attack sequences. Although demonstrated in a water-treatment SCADA environment, its graph represents paths permitted by a predefined system and vulnerability model and not behaviors reconstructed from process evidence. Cyber-Physical Attack Graphs (CPAGs)~\cite{barrere2023cyber} extend this paradigm by explicitly modeling cyber and physical actions and their propagation, supporting cyber-physical risk analysis. However, their dependencies are likewise derived from explicit system and attack knowledge. Bayes-CAPS~\cite{sahu2023inferring} introduces stronger evidence-conditioned reasoning by using Bayesian attack graphs and observations such as alerts and system evidence to update the probability of compromise states in cyber-physical power systems. While closer to the present work in its treatment of uncertainty and adversarial progression, its graph structure still originates from modeled access paths, vulnerabilities, and compromise states, without covering ATT\&CK observations inferred from numerical process telemetry.

Overall, the key distinction is between modeling how an adversary could progress and reconstructing what the available evidence suggests did occur. In this context, the proposed framework adopts the latter perspective through the construction of an evidence-driven attack dependency graph. Nodes represent time-indexed ATT\&CK observations inferred from DT state changes, while candidate relationships are evaluated using temporal compatibility, process stage, asset context, DT operational relationships, and semantic reasoning. Accepted edges distinguish \texttt{Causal}, \texttt{Temporal}, and \texttt{Co-occurrence} relationships, whereas a \texttt{None} class allows abstention when the evidence does not adequately justify a dependency. This separation between operational relatedness and adversarial causality underpins the dependency analysis presented later in the paper.

\subsection{Positioning of the Proposed Work}
\label{sec:background_positioning}

As discussed in sections~\ref{sec:background_attck} to~\ref{sec:background_graphs}, the reviewed literature provides mature capabilities within three largely distinct research directions. DT-based cybersecurity offers synchronized state representations, runtime reference models, experimentation environments, and anomaly detection; ATT\&CK mapping provides semantic translation from security-oriented text to standardized adversarial behavior; and attack-graph research models multi-stage cyber and cyber-physical compromise. The gap addressed here lies in integrating these capabilities around observed ICS process evidence. Table~\ref{tab:related_work_comparison} summarizes the literature's positioning. Specifically, the comparison does not assume that all approaches address the same problem; rather, it identifies capabilities relevant to the task considered here: synchronized cyber-physical context, direct use of operational telemetry, ATT\&CK grounding, semantic interpretation, evidence-conditioned graph construction, typed dependency semantics, and evaluation across multiple datasets or system instances. In the table, \texttt{DT} denotes whether a DT is a central system representation; \texttt{Tel.} whether operational telemetry is used directly for security inference; \texttt{ATT\&CK} whether an explicit ATT\&CK grounding exists; \texttt{Sem.} whether semantic or language-based ATT\&CK interpretation is displayed; \texttt{Graph} whether attack/dependency graph construction is explored; \texttt{Evid.} whether graph reasoning is conditioned or updated using observed evidence; \texttt{Typed} whether explicit semantics distinguish multiple forms of inter-step dependency; and \texttt{Multi} whether evaluation spans multiple datasets or system instances.

\begin{table*}[!htbp] 
\centering 
\caption{Comparison of representative work and the proposed framework. Y: explicit support; P: partial support; --: outside scope.} 
\label{tab:related_work_comparison} 
\scriptsize 
\setlength{\tabcolsep}{3.2pt} 
\resizebox{\linewidth}{!}{ 
\begin{tabular}{@{}lcccccccc@{}} 

\toprule \textbf{Year/Work} & \textbf{DT} & \textbf{Tel.} & \textbf{ATT\&CK} & \textbf{Sem.} & \textbf{Graph} & \textbf{Evid.} & \textbf{Typed} & \textbf{Multi} \\ 

\midrule 
2020~\cite{gehrmann2019digital} & Y & P & -- & -- & -- & -- & -- & -- \\ 2022~\cite{kandasamy2022electric} & Y & -- & -- & -- & -- & -- & -- & -- \\ 
2022~\cite{liu2022threat} & -- & -- & Y & Y & -- & -- & -- & -- \\ 
2023~\cite{xu2023digital} & Y & Y & -- & -- & -- & -- & -- & Y \\ 
2023~\cite{sahu2023inferring} & -- & Y & -- & -- & Y & Y & P & Y \\ 
2023~\cite{barrere2023cyber} & -- & -- & -- & -- & Y & -- & P & -- \\ 
2024~\cite{moradi2024crystal} & Y & Y & -- & -- & P & -- & -- & Y \\ 
2024~\cite{alcaraz2024digital} & Y & Y & -- & -- & -- & -- & -- & -- \\ 
2024~\cite{li2024automated} & -- & -- & Y & Y & -- & -- & -- & -- \\ 
2020~\cite{al2019a2g2v} & -- & -- & -- & -- & Y & -- & -- & Y \\ 
2025~\cite{xu2025plc} & Y & Y & -- & -- & -- & -- & -- & -- \\ 
\midrule 

\textbf{Proposed framework} & \textbf{Y} & \textbf{Y} & \textbf{Y} & \textbf{Y} & \textbf{Y} & \textbf{Y} & \textbf{Y} & \textbf{Y} \\ 

\bottomrule 
\end{tabular} 
} 
\end{table*}

Observing Table~\ref{tab:related_work_comparison}, three boundaries emerge. First, DT-assisted cybersecurity methods use synchronized or model-derived state for monitoring, anomaly detection, and experimentation, but the reviewed approaches do not translate such state changes into ATT\&CK-grounded incident representations. Second, semantic ATT\&CK methods infer adversarial concepts from security-language descriptions, but generally assume that the relevant behavior is already expressed textually and not derived from synchronized process telemetry. Third, cyber-physical attack graphs model vulnerabilities, compromise paths, physical consequences, and uncertain evidence, but are typically instantiated from predefined system and attack models instead of incrementally from ATT\&CK observations inferred from operational data. Motivated by these gaps, the proposed contribution lies in integrating these capabilities without replacing any individual component. Synchronized DT state is transformed into evidence-preserving semantic descriptions; retrieval-constrained reasoning maps those descriptions to ATT\&CK for ICS; and the resulting time-indexed observations are incrementally organized into a typed dependency graph. This integration also exposes a central analytical tension. That is, DT context can disambiguate sparse evidence and improve semantic interpretation, but the same context may make an incorrect upstream hypothesis appear operationally plausible during dependency inference. Therefore, the evaluation examines both whether DT context improves ATT\&CK mapping and under which conditions it helps or harms reconstruction of attack progression.

% ═══════════════════════════════════════════════════════════════════════════════
\section{DT-Centered System Architecture}
\label{sec:arch}

The proposed framework is centered on a DT that maintains a time-indexed representation of the monitored ICS, including asset structure, process relationships, sensor measurements, actuator and control states, nominal operating conditions, and historical observations. For offline evaluation, the DT is synchronized by replaying SWaT, WADI, and BATADAL telemetry in timestamp order; note that in an operational deployment, the same mechanism could process live ICS telemetry. The architecture addresses two representational gaps. First, numerical telemetry alone lacks the cyber-physical, operational, and temporal context required for reliable LLM reasoning. To this end, the DT contextualizes each observation with the affected asset, related process elements, prior state, and deviation from expected behavior. Second, identifying ATT\&CK techniques does not by itself capture how adversarial actions relate over time. Thus, the framework addresses this through typed dependency inference and incremental attack-graph construction.

\begin{figure}[pos=htbp]
    \centering
    \includegraphics[width=\linewidth]{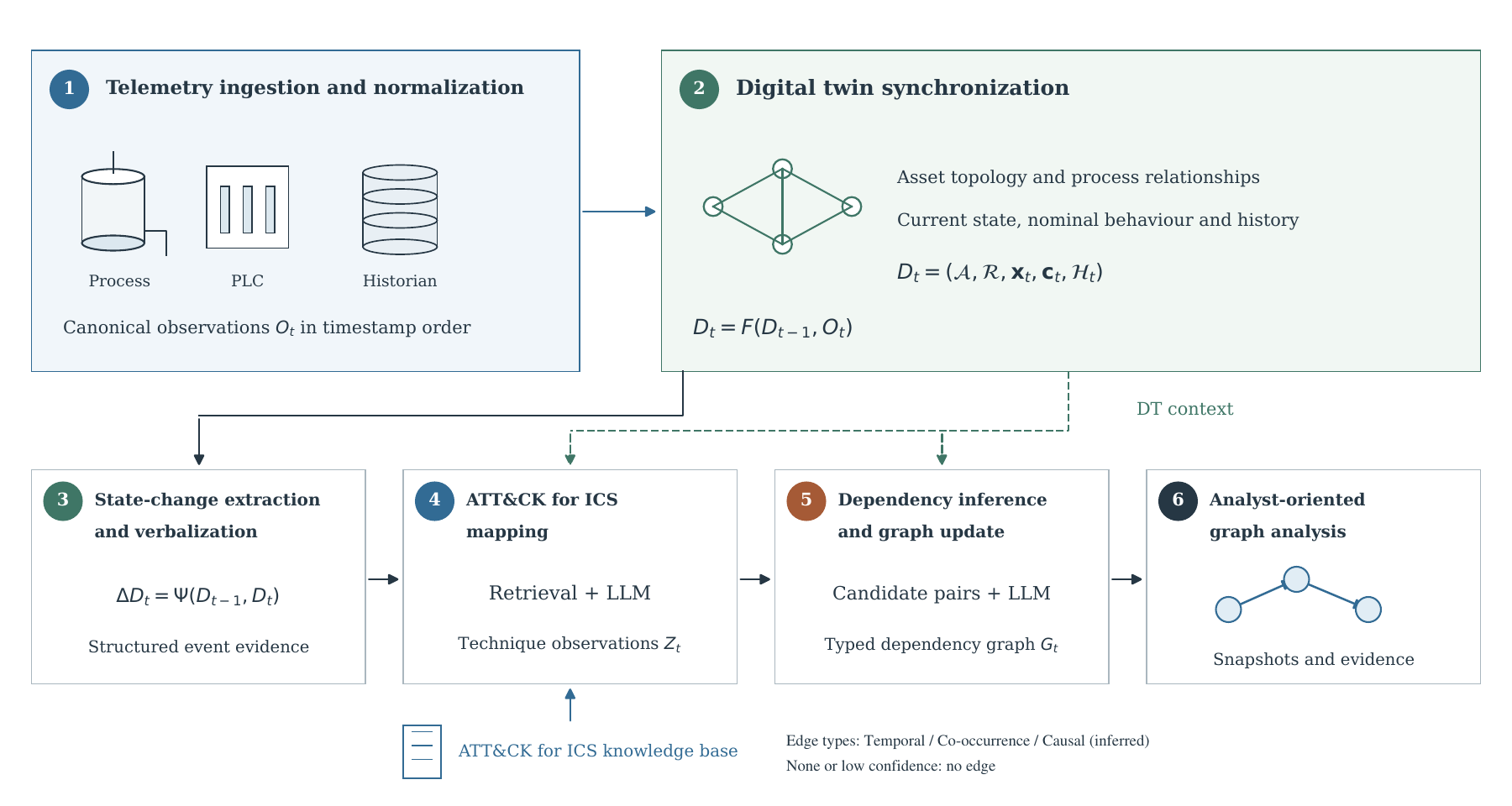}
    \caption{Overview of the six-stage DT-assisted ATT\&CK reasoning framework.}
    \label{F:arch}
\end{figure}

As illustrated in Figure~\ref{F:arch}, the framework comprises six stages. Stage~1 ingests and normalizes heterogeneous telemetry into a canonical representation. Stage~2 synchronizes the DT and updates its cyber-physical state. Stage~3 identifies meaningful state changes and converts them into structured natural-language event descriptions. Stage~4 uses a retrieval-augmented LLM, grounded in the ATT\&CK for ICS knowledge base, to map these descriptions to techniques. Stage~5 classifies candidate relationships between time-indexed technique observations as \texttt{Causal}, \texttt{Temporal}, or \texttt{Co-occurrence} and updates the dependency graph. Stage~6 analyzes the evolving graph, projects observations onto ATT\&CK tactic phases, and highlights relationships requiring further analyst investigation. Collectively, the DT serves as the framework's persistent state and contextual reasoning layer, while the LLM operates on structured descriptions of synchronized state changes. For the readers' reference, dataset attack labels are withheld during inference and used only for experimental evaluation.

\subsection{Formal Model}
\label{S:formal}

\paragraph{DT state.} Let $\mathcal{O}_t$ denote the set of telemetry observations received at time $t$. The DT state is defined as:

$$
D_t=(\mathcal{A},\mathcal{R},\mathbf{x}_t,\mathbf{c}_t,\mathcal{H}_t),
\qquad
D_t=F(D_{t-1},\mathcal{O}_t),
$$

where $\mathcal{A}$ is the set of modeled assets, $\mathcal{R}$ is the set of structural and process relationships among those assets, $\mathbf{x}_t$ represents the physical process state, $\mathbf{c}_t$ represents the cyber and control state, and $\mathcal{H}_t$ contains the retained history of previous DT states. For the benchmark datasets, the synchronization function is executed by replaying observations in timestamp order. Attack labels are excluded from $\mathcal{O}_t$ during inference.

\paragraph{Twin-state change representation.} The framework derives a contextual state-change representation from successive DT states as:

$$
\Delta D_t=\Psi(D_{t-1},D_t),
$$

where $\Psi$ identifies meaningful cyber-physical changes according to the frozen state-change criteria used in the implementation, including discrete state transitions, deviations from nominal operating ranges, unusually large rates of change, persistent abnormal conditions, and abnormal hold behaviour. Changes that do not satisfy the configured significance criteria are not forwarded to the LLM reasoning stage.

\paragraph{ATT\&CK technique observations.} Let $\mathcal{K}$ denote the ATT\&CK for ICS knowledge base and $\mathcal{T}$ the set of techniques contained in $\mathcal{K}$. Given DT state changes over a window of length $w$, the mapping function produces:

$$
\mathcal{Z}_t=M(\Delta D_{t-w:t},\mathcal{K}),
\qquad
z_i=(\tau_i,t_i,a_i,q_i,\epsilon_i),
$$

where $\tau_i\in\mathcal{T}$ is an ATT\&CK for ICS technique, $t_i$ is its observation time, $a_i\in\mathcal{A}$ is the associated asset or process component, $q_i\in[0,1]$ is the confidence score returned by the mapping stage, and $\epsilon_i$ is the supporting evidence extracted from the verbalized DT state changes.

\paragraph{Evolving attack dependency graph.} At time $t$, the framework maintains a directed, typed, and weighted attack dependency graph:

$$
G_t=(V_t,E_t,\lambda,\delta,\omega),
$$

where each node $v_i\in V_t$ represents a time-indexed technique observation and not a unique ATT\&CK technique identifier. Consequently, the same ATT\&CK technique may appear in multiple nodes when observed at different times, on different assets, or during different attack stages. The set $E_t\subseteq V_t\times V_t$ contains the inferred dependency edges. The graph functions are defined compactly as:

\[
\begin{gathered}
\lambda:V_t\rightarrow\Phi,\qquad
\delta:E_t\rightarrow\Gamma,\qquad
\omega:E_t\rightarrow[0,1],\\
\Gamma=\{\texttt{Causal},\texttt{Temporal},\texttt{Co-occurrence}\}.
\end{gathered}
\]

Here, $\lambda$ assigns each node its ATT\&CK tactic phase, $\delta$ assigns each edge its relation type, and $\omega$ assigns the confidence score returned by the dependency-inference stage. This score represents the confidence of the LLM inference and should not be interpreted as a calibrated posterior probability. The graph is updated incrementally according to:

$$
G_t=U(G_{t-1},\mathcal{Z}_t,\Delta D_{t-w:t}),
$$

where $U$ adds new technique-observation nodes, selects candidate node pairs, infers dependency relationships, and updates existing edge evidence.

% ═══════════════════════════════════════════════════════════════════════════════
\subsection{Operational Pipeline}
\label{sec:pipeline}

\paragraph{Stage 1: Telemetry ingestion and canonicalization.}

Telemetry from SWaT, WADI, and BATADAL is ingested through dataset-specific adapters and transformed into a common observation format. The adapters standardize timestamps, asset identifiers, variable names, data types, sampling intervals, and measurement units while preserving the original temporal ordering of the observations. The canonical observation schema is defined as follows:

\begin{center}
\small
\texttt{[timestamp, asset\_id, asset\_type, process\_stage, variable, value, previous\_value, sampling\_interval, attack\_label]}
\end{center}

The timestamp field records the observation time, asset\_id uniquely identifies the associated sensor, actuator, controller, tank, pump, valve, or hydraulic component, and process\_stage identifies its location within the monitored process. The previous\_value field supports state-change analysis, while sampling\_interval preserves differences in temporal resolution across the datasets. Dataset-specific attack labels are retained in a separate evaluation layer and are not supplied to the DT, retriever, or LLM during inference. A second metadata structure describes the static cyber-physical context associated with each modeled asset:

\begin{center}
\small
\texttt{[asset\_id, asset\_type, process\_stage, variable\_role, nominal\_range, control\_role, upstream\_assets, downstream\_assets]}
\end{center}

The asset and relationship metadata is constructed from dataset documentation, process diagrams, and variable descriptions. Attack-scenario descriptions are used exclusively to construct evaluation annotations and are not supplied to DT construction, retrieval, or LLM inference. Nominal ranges and expected transition patterns are estimated only from the designated normal dataset or the normal portion of the training partition. Notably, test attack intervals are not used to determine nominal behavior. Missing metadata is explicitly represented as unknown without being inferred without supporting evidence. Subsequently, the canonicalization stage does not remove anomalous observations or balance the data because the temporal continuity of each attack scenario must be preserved. Discrete actuator states are carried forward until a new state is observed, while continuous measurements retain their recorded values and timestamps. Any additional imputation or interpolation is performed only when required by the dataset and is documented separately for each dataset.

\paragraph{Stage 2: DT state synchronization.}

The canonical observations are used to instantiate and synchronize a state-based DT of each monitored system. The twin contains a static asset model, a process-relationship model, the current physical state, the current cyber and control state, and a retained history of previous states, as defined in section~\ref{S:formal}. Recall also from section~\ref{S:formal} that the static component of the twin contains the asset set $\mathcal{A}$ and the relationship set $\mathcal{R}$. Relationships may represent physical flow, process-stage membership, upstream and downstream dependencies, control relationships, or shared operational functions. The dynamic component contains the physical process state $\mathbf{x}_t$, the cyber and control state $\mathbf{c}_t$, and the historical state sequence $\mathcal{H}_t$. Control-state fields are populated only when the corresponding actuator, command, or controller information is available in the dataset.

For SWaT, the twin represents the six treatment stages and the relationships among tanks, pumps, valves, flow sensors, pressure sensors, and level sensors. For WADI, it represents the assets and process relationships of the water distribution subsystems. For BATADAL, it represents the hydraulic components, measured variables, and connectivity information available for the simulated C-Town water network. Dataset-specific details are retained within the adapters while exposing a common interface to the subsequent stages. Each synchronized state retains both the current observation and the relevant operational context. Consequently, a pump transition is represented together with its previous state, associated process stage, related sensor values, nearby valve states, expected operating conditions, and recent history. This context allows the framework to distinguish an isolated value change from a potentially meaningful cyber-physical event.

\paragraph{Stage 3: Twin-state change extraction and verbalization.}

Processing every synchronized value would generate redundant event descriptions and excessive LLM input. Therefore, the framework applies the state-change function $\Psi$, as defined in section~\ref{S:formal}, to successive DT states. Recall also from section~\ref{S:formal} that the function $\Psi$ identifies telemetry changes that satisfy the frozen state-change criteria used in the implementation. Thresholds are derived from normal-operation data and remain frozen during attack evaluation. Operational relationships among assets are not themselves used as evidence that an attack occurred; instead, the corresponding DT context is attached to detected state changes during verbalization and is made available to the later semantic reasoning stages.

\begin{itemize}
    \item \textbf{Discrete transition:} An actuator, pump, valve, controller, or operating-mode variable changes from one discrete state to another.
    \item \textbf{Nominal-range deviation:} A continuous variable moves outside its normal operating range.
    \item \textbf{Rate-of-change deviation:} The change in a continuous variable exceeds an asset-specific threshold $\eta_j$.
    \item \textbf{Persistent deviation:} An abnormal condition remains present for at least $p$ consecutive observations.
\end{itemize}

The thresholds $\eta_j$, the persistence parameter $p$, and the sliding-window length $w$ are selected using the normal training data or a separate validation partition. They are not optimized using the test attack scenarios. Significant changes occurring within a sliding window of length $w$ are grouped according to their asset, process stage, and temporal proximity. The grouping process reduces repeated descriptions of the same persistent condition while preserving transitions that occur on different assets or at different attack stages. Each grouped state change is converted into a structured natural-language event description. The verbalization records timestamp, asset/type, process stage, previous/current state, nominal range, related-asset state, persistence, and recent history.

The template is populated only with context supported by the DT. Fields that are unavailable for a specific asset or dataset are marked as unknown or omitted. The verbalization stage does not describe an event as malicious and does not expose attack labels to the LLM. Its purpose is to express the observed cyber-physical state change without assigning an ATT\&CK interpretation in advance. The resulting event description retains the evidence required for later traceability, including the source timestamps, affected assets, measurements, previous states, and relevant process relationships. This information allows each ATT\&CK mapping to be traced back to the DT observations that produced it.

\paragraph{Stage 4: RAG-based ATT\&CK mapping.}

The verbalized DT state changes are mapped to ATT\&CK for ICS techniques using Retrieval-Augmented Generation (RAG). The retrieval corpus is constructed from the ATT\&CK for ICS knowledge base and contains the identifier, name, description, associated tactics, relevant platforms, data sources, detection guidance, and documented relationships of each technique. For each verbalized event window, the retriever returns the top-$k$ most relevant ATT\&CK for ICS technique descriptions. The retrieval query combines the event description with the affected asset types, process-stage information, observed state transitions, and available cyber-physical relationships. This prevents retrieval from depending only on generic terms such as pump, valve, sensor, or anomaly. The model may return \texttt{NoMatch}, avoiding forced mappings for ambiguous process deviations. The tactic phase and official technique name are attached programmatically from the knowledge base after the \texttt{TechniqueID} has been validated. Any identifier that does not exist in the retrieved knowledge base is rejected. Each accepted technique observation is represented as $z_i$, defined in section~\ref{S:formal}. With reference to section~\ref{S:formal}, technique observations with confidence below the mapping threshold $\theta_{\mathrm{map}}$ are discarded. The retrieval depth and mapping threshold are frozen before the held-out and cross-dataset evaluations at $k=8$ and $\theta_{\mathrm{map}}=0.50$. The same values are retained for SWaT, BATADAL, and WADI and are not returned based on external-test outcomes. The implementation uses structured LLM outputs and caches completed first-run responses. No explicit temperature parameter is configured; consequently, the reported results correspond to the frozen evaluated outputs and do not assume perfectly deterministic generation.

\paragraph{Stage 5: Context-guided dependency inference and graph update.}

The accepted technique observations are used to infer \texttt{temporal}, \texttt{causal}, and \texttt{co-occurrence} relationships. Recall that the framework does not compare every possible pair of observations because exhaustive pairwise comparison scales quadratically and introduces many implausible candidates. Instead, the DT context is used to construct a restricted candidate-pair set. Two observations $(z_i,z_j)$ become a candidate pair when $t_i \leq t_j$, their temporal separation does not exceed the maximum dependency interval $\Delta_{\max}$, and at least one contextual relationship is present. Contextual relationships include involvement of the same asset, involvement of assets within the same process stage, an upstream or downstream process dependency, a shared control relationship, or a state change in one asset followed by a physically consistent change in another asset. Accordingly, the candidate-pair set is represented as:

\[
\mathcal{C}_t =
\left\{
(z_i,z_j)
\;\middle|\;
0 \leq t_j-t_i \leq \Delta_{\max}
\land
\chi(z_i,z_j)=1
\right\},
\]

where $\chi(z_i,z_j)$ equals one when at least one DT context condition is satisfied. For each candidate pair, the LLM receives the two ATT\&CK technique definitions, their timestamps, associated assets, evidence spans, relevant DT states, and the process relationship between the affected assets. The dependency classifier receives both observations, their ATT\&CK definitions, supporting evidence, timestamps, assets, and supplied DT context, and returns one of \texttt{Causal}, \texttt{Temporal}, \texttt{Co-occurrence}, or \texttt{None} with confidence and justification.

The inclusion of \texttt{None} prevents the model from being forced to create an edge for every candidate pair. Relationships classified as None or assigned a confidence below the edge threshold $\theta_{\mathrm{edge}}$ are not added to the graph. Each accepted edge is represented as:

\[
e_{ij} = (v_i,v_j,\delta_{ij},\omega_{ij},\epsilon_{ij}),
\]

where $v_i$ and $v_j$ are time-indexed technique-observation nodes, $\delta_{ij} \in \{\texttt{Causal},\texttt{Temporal},\texttt{Co-occurrence}\}$ is the inferred relation type, $\omega_{ij} \in [0,1]$ is the inference confidence, and $\epsilon_{ij}$ records the supporting evidence and justification. A \texttt{Causal} edge indicates an inferred enabling relationship supported by the available evidence and should not be interpreted as definitive proof of causation. A Temporal edge indicates ordered observations without sufficient causal evidence. A Co-occurrence relationship does not impose an attack order and is excluded from time-respecting attack-path extraction. In the graph implementation, co-occurrence may be stored as paired reciprocal edges with the same confidence value. The evolving attack dependency graph is updated according to the $G_t$ function defined in section~\ref{S:formal}.

\paragraph{Stage 6: Analyst-oriented graph analysis.}

The evolving dependency graph provides an analyst-oriented representation of how ATT\&CK hypotheses accumulate and become related over time. This stage does not introduce additional attack labels or claim to recover a definitive adversary path. Instead, it derives descriptive views from the inferred graph. Graph snapshots preserve the state of the dependency representation at successive update times and expose when new ATT\&CK observations and relationships become available. A tactic-level projection can additionally summarize which ATT\&CK for ICS tactic phases are represented by the current hypotheses. Co-occurrence relationships are retained as unordered associations, while Temporal and Causal predictions preserve their inferred direction. These outputs are intended to support investigation and qualitative progression analysis. Because the available benchmark documentation does not provide complete path-level ground truth, full attack-path reconstruction and graph-centrality accuracy are not treated as independently validated contributions in this study.

\section{Experimental Evaluation}
\label{sec:eval}
% ═══════════════════════════════════════════════════════════════════════════════

\subsection{Experimental Setup}

The evaluation is organized into three complementary components. SWaT serves as the primary benchmark for the complete mapping ablation and the dependency experiments. BATADAL and WADI are used as external ATT\&CK mapping-validation datasets. Dependency performance is not reported for BATADAL or WADI because their available scenario documentation does not provide sufficiently complete action-level dependency ground truth to support a defensible edge evaluation. For SWaT, the DT maintains metadata for all 51 process and actuator tags, including process stage, asset type, variable role, control role, and conservative operational upstream/downstream relationships. Continuous-value nominal ranges are estimated exclusively from normal-operation data using the 0.5th and 99.5th percentiles. The DT event detector uses frozen z-score and change thresholds of 4, a persistence requirement of three samples, and normal-derived hold thresholds. Significant state changes are grouped into 60~s evidence windows. Technique observations with the same technique, process stage, and asset set are consolidated into episodes when their temporal separation does not exceed 900~s.

BATADAL is processed using its normal-operation trace to construct a dataset-specific DT nominal profile and its seven documented attack scenarios for external mapping evaluation. Because BATADAL is sampled hourly, state changes are grouped into 3600~s semantic windows, and the episode-consolidation gap is set to 54,000~s, corresponding to the same pre-specified fifteen-window temporal scale used for the one-second datasets. These are sampling-rate adaptations and are not selected from attack-performance results. WADI is processed from the corrected WADI.A2 traces. Timestamps are reconstructed from the preserved one-second row ordering and the documented dataset start time. A removed interval in the normal trace is retained as a temporal gap and is not interpolated. The normal profile contains 123 modeled variables with usable baseline observations; four additional metadata variables contain no valid normal-operation samples and are retained in the asset metadata but excluded from data-driven nominal-state estimation. WADI state changes are grouped into 60~s windows and consolidated with the frozen 900~s episode gap. A 3600~s suppressed warm-up period is used before each evaluated scenario, while pre-attack context is not exported to the mapping stage.

The ATT\&CK retrieval and mapping configuration is frozen across all three datasets. The knowledge base is ATT\&CK for ICS STIX release v19.1, from which 97 techniques are indexed. Retrieval combines dense similarity from sentence-transformers/all-mpnet-base-v2 with lexical TF--IDF similarity using weights of 0.70 and 0.30, respectively. The top $k=8$ candidates are supplied to retrieval-augmented variants, and the mapping confidence threshold is fixed at $\theta_{\mathrm{map}}=0.50$. No retrieval weight, candidate count, confidence threshold, or prompt is retuned after observing BATADAL or WADI results. The experiments were conducted between May and September 2026 using the API model identifier \texttt{gpt-5-mini}. Structured JSON outputs are validated against fixed schemas. For the primary ablation, external-validation, and dependency experiments, the first completed responses are cached and retained as the frozen primary outputs. The implementation does not explicitly configure a temperature parameter, and no assumption of perfectly deterministic generation is made.

To assess generation-level variability of the principal SWaT comparison, the M3 and M4 mapping configurations are independently executed two additional times, yielding three complete runs in total. Each repeat uses a fresh isolated LLM cache while preserving the same held-out scenarios, DT event windows, ATT\&CK knowledge base, retrieval configuration, prompts, mapping threshold, and 900~s episode-consolidation rule. No parameter is retuned between runs. M1, M2, BATADAL, WADI, and the dependency experiments retain their frozen primary outputs and are not included in the repeated-generation analysis. Table~\ref{tab:experimental_parameters} summarizes the principal frozen parameters used throughout the evaluation. Dataset-specific differences are limited to temporal scaling required by the underlying sampling cadence; the retrieval, mapping, and confidence settings remain unchanged across datasets.

\begin{table}[!htbp]
\centering
\caption{Principal experimental parameters. Parameters marked as asset-specific are derived exclusively from normal-operation data. BATADAL temporal parameters are scaled to its one-hour sampling cadence.}
\label{tab:experimental_parameters}
\fontsize{9}{11}\selectfont
\setlength{\tabcolsep}{3pt}
\renewcommand{\arraystretch}{1.15}
\begin{tabularx}{\linewidth}{@{}>{\raggedright\arraybackslash}p{0.15\linewidth}>{\raggedright\arraybackslash}p{0.18\linewidth}ccc>{\raggedright\arraybackslash}X@{}}
\toprule
\textbf{Component} &
\textbf{Parameter} &
\textbf{SWaT} &
\textbf{WADI} &
\textbf{BATADAL} &
\textbf{Selection basis} \\
\midrule

DT profile
& Nominal quantiles
& $[0.005,0.995]$
& $[0.005,0.995]$
& $[0.005,0.995]$
& Normal-operation data only \\

DT extraction
& State z-score threshold
& $4$
& $4$
& $4$
& Frozen before final evaluation \\

DT extraction
& Rate/change z-score threshold
& $4$
& $4$
& $4$
& Frozen before final evaluation \\

DT extraction
& Persistence requirement
& $3$ samples
& $3$ samples
& $3$ samples
& Frozen before final evaluation \\

DT extraction
& Continuous hold threshold
& \multicolumn{3}{c}{$\max(\mathrm{normal\ hold})+1$}
& Asset-specific, normal data only \\

Evidence grouping
& Semantic window
& $60$ s
& $60$ s
& $3600$ s
& One analysis window per dataset cadence \\

Episode construction
& Maximum merge gap
& $900$ s
& $900$ s
& $54{,}000$ s
& Fifteen semantic windows \\

Retrieval
& Dense / lexical weights
& $0.70/0.30$
& $0.70/0.30$
& $0.70/0.30$
& Frozen hybrid retriever \\

Retrieval
& Candidate depth $k$
& $8$
& $8$
& $8$
& Frozen before held-out evaluation \\

Mapping
& Confidence threshold $\theta_{\mathrm{map}}$
& $0.50$
& $0.50$
& $0.50$
& Frozen before held-out evaluation \\

Dependency
& Candidate temporal gap $\Delta_{\max}$
& $900$ s
& N/A
& N/A
& SWaT dependency evaluation only \\

Dependency
& Edge threshold $\theta_{\mathrm{edge}}$
& $0.50$
& N/A
& N/A
& Same confidence acceptance rule \\

LLM
& Backend
& \texttt{gpt-5-mini}
& \texttt{gpt-5-mini}
& \texttt{gpt-5-mini}
& Fixed across experiments \\

\bottomrule
\end{tabularx}
\end{table}

The frozen values are retained for all primary results even when the subsequent sensitivity analysis identifies settings with numerically higher held-out performance; the sensitivity study is used only to assess robustness and does not trigger parameter reselection. Two SWaT scenarios, A21 and A22, are used only during development for prompt construction, retrieval inspection, and dependency-prompt sanity testing. The final SWaT mapping benchmark contains nine held-out scenarios: A03, A08, A10, A16, A26, A31, A37, A39, and A41. Five previously unused scenarios, A23, A25, A27, A30, and A35, form the positive coordination benchmark. The exploratory end-to-end dependency experiment reuses the frozen M4 mapping outputs from the nine held-out scenarios. All seven documented BATADAL attacks and all fourteen corrected WADI temporal attack scenarios are retained for external mapping validation, including scenarios with no telemetry-evaluable primary ATT\&CK node so that unsupported predictions remain observable as False Positives (FPs). Scenario membership and exclusion criteria were fixed before evaluation of the final configurations. No evaluated scenario was subsequently removed or reassigned on the basis of mapping or dependency results.

\subsection{Ground-Truth Construction}
\label{sec:ground_truth}

ATT\&CK ground truth is constructed from documented scenario semantics and not from binary attack labels. Three concepts are separated throughout the annotation process: scenario truth, which records the adversarial behavior described by the scenario documentation; telemetry observability, which records whether the corresponding action or effect is represented in the available historian; and mapping evidence level, which determines whether the observed telemetry directly represents the relevant action artifact or supports only an inferred interpretation. This separation prevents physical consequences of an attack from being treated automatically as direct evidence of the attack mechanism. Two authors independently constructed the ground-truth annotations, including scenario truth, telemetry observability, mapping evidence level, and dependency relationships. Disagreements were resolved through discussion to produce the final annotations used for evaluation.

For SWaT, published scenario descriptions identify the manipulated process points and documented attacker behavior. Unauthorized command actions are annotated as T1692.001 (Command Message) when supported by the scenario description. A device state alone, however, does not reveal the provenance of the command that produced that state. Command-message nodes for which the malicious command artifact is unavailable are retained as scenario truth but excluded from quantitative mapping recall. Manipulated sensor reports are annotated as T1692.002 (Reporting Message) when the falsified process value is visible in the historian. These nodes may be telemetry-evaluable even when the underlying network message is not directly recorded.

BATADAL attack intervals are reconstructed from the original scenario descriptions and not from the intentionally incomplete binary labels. Six primary telemetry-evaluable nodes are used for quantitative evaluation, all corresponding to T1692.002 reporting manipulation. Documented pump-speed modification remains part of scenario truth but is not used as a quantitative target when the relevant speed-setpoint artifact is absent from the recorded telemetry. One documented attack scenario hence contains no primary quantitative ATT\&CK node and acts as an attack-context negative control: predictions within that interval may contribute FPs, but the scenario does not contribute to recall or macro-F1.

For WADI, the corrected attack table defines the scenario intervals and semantics, while the supplied binary label is retained only as a consistency check. Device \texttt{\_STATUS} variables are not interpreted as direct evidence of an unauthorized command because they expose resulting device state without command provenance. In contrast, modulating-valve \texttt{\_CO} channels are treated as direct control-output artifacts when the documented attack explicitly targets the corresponding control output. The recorded \texttt{2\_PIC\_003\_SP} channel similarly provides direct evidence for the documented parameter modification in Attack~13. Manipulated sensor-value scenarios are represented as T1692.002 with inferred evidence when the falsified reported value is visible. The resulting WADI quantitative ground truth contains eight primary nodes: four T1692.001 Command Message nodes, three T1692.002 Reporting Message nodes, and one T0836 Modify Parameter node. Additional documented but non-observable actions remain part of scenario truth without entering the recall denominator.

For episode-level mapping evaluation, a predicted episode is matched to at most one primary ground-truth node. A match requires the same ATT\&CK technique, the same process stage, at least one overlapping asset, and temporal overlap between the predicted and annotated intervals. When multiple ground-truth nodes satisfy these conditions, the node with the largest asset overlap is selected.

Dependency ground truth is intentionally more conservative. The five SWaT coordination scenarios contain 12 documented action nodes and nine annotated action pairs. The documentation establishes that the actions form coordinated multi-point attacks but does not provide sufficient evidence for a reliable internal ordering or directional causal relation. Consequently, all nine positive relations are labeled Co-occurrence. No Causal or Temporal relation is manufactured solely from historian timing.

\subsection{Ablation and Baseline Configurations}
\label{sec:ablations}

Four mapping configurations are evaluated to isolate the contribution of DT-derived context and ATT\&CK retrieval:

\begin{itemize}
    \item \textbf{M1: Structured evidence without RAG.} The LLM receives structured event-level evidence without DT-enriched semantic context or ATT\&CK retrieval.
    \item \textbf{M2: Verbalized evidence without RAG.} DT events are converted into natural-language descriptions before LLM mapping, but the full ATT\&CK catalog is not retrieved per observation.
    \item \textbf{M3: RAG without DT structural context.} Evidence windows are mapped using ATT\&CK retrieval, but process-stage topology, control roles, and upstream/downstream DT context are omitted.
    \item \textbf{M4: DT context with RAG.} The complete proposed mapping pipeline combines DT state-change evidence, asset and process context, and ATT\&CK retrieval.
\end{itemize}

M3 and M4 use identical evidence windows, retrieved candidate lists, mapping thresholds, and episode-consolidation rules. Their difference is restricted to the structural DT context supplied to the mapping LLM: M4 additionally receives process relationships, control roles, and upstream/downstream context. Therefore, this comparison evaluates the incremental contribution of structural context during semantic mapping, conditional on the shared DT-derived evidence and retrieval outputs.

Dependency inference is evaluated using four analogous configurations:

\begin{itemize}
    \item \textbf{D1: Time-only heuristic.} Overlapping intervals are labeled Co-occurrence, non-overlapping pairs within 900\,s are labeled Temporal, and all other pairs are labeled None.
    \item \textbf{D2: Deterministic DT rules.} Temporal compatibility is combined with same-stage membership, shared assets, or direct DT topology links.
    \item \textbf{D3: LLM without DT structural context.} The LLM receives the ATT\&CK episodes, asset identifiers, evidence levels, and timestamps, but no process-stage semantics, control roles, or topology.
    \item \textbf{D4: DT-aware LLM.} The proposed method additionally receives process stage, asset type, variable and control roles, upstream/downstream asset context, and direct operational topology links.
\end{itemize}

D1--D4 use the same frozen temporal gap and confidence settings across the dependency experiments. No parameter is retuned after observing the final benchmark outputs.

\subsection{Evaluation Metrics}
\label{sec:evaluation_metrics}

\paragraph{ATT\&CK Mapping.}
Episode-level micro Precision, Recall, and F1-score are computed from one-to-one matches between predicted episodes and primary telemetry-evaluable ground-truth nodes. Macro F1 is additionally reported over scenarios containing at least one primary ground-truth node. Scenarios with zero primary nodes remain in the experiment and all predictions made in those intervals contribute FPs, but such scenarios are excluded from the macro-F1 denominator. The matching criterion is intentionally strict. A predicted episode is considered correct only when the ATT\&CK technique, process stage, asset evidence, and temporal interval agree with a telemetry-evaluable ground-truth node. Consequently, an observation may correspond to genuine abnormal process behavior while still being counted as a FP ATT\&CK attribution if the available evidence does not support the documented adversarial mechanism. Therefore, the resulting scores evaluate semantic attack attribution, not binary anomaly detection.

FP episodes are reported explicitly because unsupported ATT\&CK hypotheses become candidate nodes for subsequent dependency inference. The full M1--M4 ablation and paired statistical analysis are performed on the nine held-out SWaT scenarios. BATADAL and WADI are treated as external validation datasets and compare only the frozen M3 without structural DT context configuration against the complete DT-enriched M4 configuration.

Paired per-scenario F1 differences between M4 and the other mapping configurations are evaluated using the Wilcoxon signed-rank test on the frozen primary SWaT outputs. The nine held-out scenarios provide limited statistical power. Failure to reach statistical significance does not establish equivalence or absence of an effect. Variation across the three independent M3/M4 generations is reported separately and summarized descriptively.

Throughout the quantitative evaluation, $FP$ denotes a predicted episode that cannot be matched to a primary node in the frozen ground-truth annotation. Because the annotation is intentionally restricted to documented, telemetry-evaluable adversarial behavior, these values should be interpreted as annotation-relative FPs and not as proof that every unmatched semantic hypothesis is universally incorrect.

\paragraph{Positive Coordination Benchmark.}
The first dependency experiment contains nine positive Co-occurrence pairs and no negative pairs. Pair accuracy and coordination recall are hence reported together with the abstention rate. Precision and F1 are not interpreted for this benchmark because the absence of negative examples prevents estimation of FP Co-occurrence predictions.

\paragraph{End-to-End Dependency Specificity.}
The second dependency experiment evaluates whether an inferred edge should exist between predicted ATT\&CK episodes. Candidate pairs are derived from the frozen M4 mapping outputs using the same 900\,s maximum gap. Under the frozen primary ATT\&CK annotations, a pair is positive only when both predicted endpoints map to telemetry-evaluable ground-truth nodes connected by an annotated dependency edge. Otherwise, the pair is treated as an \emph{annotation-relative negative control}. These negatives indicate that no edge is supported by the annotated attack graph; they do not assert universal physical independence between the corresponding process components.

A total of 381 candidate episode pairs are generated. All observable positive pairs are retained, while up to three negative-control pairs per scenario are selected deterministically using structural strata and a SHA-256 ordering independent of D1--D4 outputs. The resulting benchmark contains 26 pairs: one observable positive edge and 25 negative controls. Because of this extreme class imbalance, Specificity and FPR are the primary edge metrics.

Edge Precision, Recall, F1, Accuracy, and Balanced Accuracy are also reported, but positive-edge Recall and F1 have high uncertainty because only one observable positive edge is available.

The end-to-end metrics evaluate edge existence independently of relation type. A predicted edge between two correctly matched endpoints is counted as a true positive when the annotation contains a relationship between those endpoints, even if the predicted relation type differs. Consequently, edge recall does not measure relation-label accuracy. The A39 case study illustrates this distinction: D4 recovers edge existence but predicts Causal for a relationship annotated as Co-occurrence.

Graph Edit Distance and full attack-path recovery are not used as primary quantitative metrics because the available SWaT documentation does not provide a sufficiently complete, directionally annotated multi-relation attack graph to support those measures without introducing inferred ground truth.

\subsection{ATT\&CK Mapping Results}
\label{sec:mapping_results}

Table~\ref{tab:mapping_ablation} reports the frozen primary-run SWaT mapping ablation across M1--M4. The nine scenarios contain ten telemetry-evaluable primary ground-truth ATT\&CK episodes. In this primary run, M4 produces the strongest aggregate mapping result, with a micro F1 of 0.1687 and macro F1 of 0.2501. Its micro Precision nevertheless remains low at 0.0959, indicating that semantic attribution from process-historian evidence remains substantially more difficult than detecting that the underlying process is abnormal. Generation-level stability of the M3--M4 comparison is evaluated separately in the generation-variability analysis below.

\begin{table}[!htbp]
\centering
\caption{Frozen primary-run ATT\&CK mapping ablation across nine held-out SWaT scenarios.}
\label{tab:mapping_ablation}
\begin{tabular}{@{}lrrrrrr@{}}
\toprule
\textbf{Method} &
\textbf{TP} &
\textbf{FP} &
\textbf{FN} &
\textbf{Prec.} &
\textbf{Rec.} &
\textbf{Micro F1} \\
\midrule
M1 & 5 & 62  & 5 & 0.0746 & 0.5000 & 0.1299 \\
M2 & 6 & 73  & 4 & 0.0759 & 0.6000 & 0.1348 \\
M3 & 7 & 119 & 3 & 0.0556 & 0.7000 & 0.1029 \\
\textbf{M4} & \textbf{7} & \textbf{66} & \textbf{3}
& \textbf{0.0959} & \textbf{0.7000} & \textbf{0.1687} \\
\bottomrule
\end{tabular}

\vspace{0.5em}
\begin{tabular}{@{}lrr@{}}
\toprule
\textbf{Method} & \textbf{Macro F1} & \textbf{Scenarios with TP} \\
\midrule
M1 & 0.1564 & 5/9 \\
M2 & 0.1218 & 5/9 \\
M3 & 0.1040 & 6/9 \\
\textbf{M4} & \textbf{0.2501} & \textbf{6/9} \\
\bottomrule
\end{tabular}
\end{table}

The absolute values in Table~\ref{tab:mapping_ablation} should be interpreted in the context of the task being evaluated. The objective is not binary anomaly detection, but episode-level attribution of sparse adversarial behaviors from process telemetry to specific ATT\&CK for ICS techniques. A prediction is counted as correct only when the technique, process stage, asset evidence, and temporal interval agree with a telemetry-evaluable ground-truth node. As a result, process behavior may be clearly abnormal while still constituting an incorrect ATT\&CK attribution if the evidence does not support the documented adversarial mechanism. These scores are not directly comparable with binary anomaly-detection results commonly reported for SWaT, WADI, or BATADAL, where the evaluation is limited to distinguishing normal from abnormal operation.

Within the frozen primary run, the principal difference between M3 and M4 is FP suppression, not increased recall. Both variants recover seven of the ten primary ground-truth episodes, corresponding to a recall of 0.70. M3 produces 119 annotation-relative FP episodes compared with 66 for M4, a reduction of approximately 44.5\%. This increases micro F1 from 0.1029 to 0.1687, while macro F1 increases from 0.1040 to 0.2501. These values characterize the frozen primary realization; the subsequent independent-generation comparison examines whether the same behavior persists across independent LLM generations.

Performance varies substantially across scenarios, including several attacks with zero F1; the aggregate improvement should therefore not be interpreted as uniformly strong ATT\&CK attribution.

Primary-run paired per-scenario comparisons favor M4 descriptively, but none of the Wilcoxon tests reaches $p<0.05$ (M4--M1: 0.0625; M4--M2: 0.1094; M4--M3: 0.0781).

\subsection{LLM Generation Variability}
\label{sec:repeatability}

The primary mapping experiment uses the first completed and subsequently frozen LLM outputs. Because the backend is generative and no explicit temperature parameter is configured, two additional independent executions of the M3 and M4 SWaT configurations are performed to examine whether the principal M3--M4 behavior persists across generations. Each execution uses a fresh isolated cache while preserving all upstream evidence, retrieval settings, prompts, thresholds, ground truth, and episode-consolidation rules.

Table~\ref{tab:mapping-repeatability} reports the three complete runs. The exact true-positive assignments vary across generations. M3 obtains recall values of 0.70, 0.70, and 0.50, while M4 obtains 0.70, 0.50, and 0.70. Consequently, the primary-run observation that M4 preserves recall at 0.70 is not itself generation invariant.

A more stable effect is observed in annotation-relative FP suppression. M4 produces fewer FP episodes than M3 in every run: 66 versus 119 in Run~1, 81 versus 113 in Run~2, and 69 versus 98 in Run~3. These correspond to relative reductions of 44.5\%, 28.3\%, and 29.6\%, respectively, with a mean per-run reduction of 34.1\%. Across the three runs, both methods obtain the same mean recall of 0.633. Mean micro F1 is 0.0997 for M3 and 0.1452 for M4, although M4 does not exceed M3 in every individual realization.

Across these three generations, M4 produces fewer annotation-relative FP episodes in every run, but its recall and F1 advantage are not consistent across runs. Recall ranges from 0.50 to 0.70 for both configurations; M4 recall is lower than M3 in Run 2 and higher in Run 3. Equal mean recall does not imply preservation of individual mappings. The reported 34.1\% reduction is the arithmetic mean of the three per-run percentage reductions, and not the percentage reduction calculated from the mean FP counts. Three generations provide a limited descriptive assessment of variability and do not establish the frequency with which the observed effect would persist across further executions.

\begin{table}[t]
\centering
\caption{SWaT mapping results across three independent LLM generations. Run 1 is the frozen primary evaluation.}
\label{tab:mapping-repeatability}
\small
\begin{tabular}{llrrrrr}
\hline
Run & Method & TP & FP & Recall & Micro F1 & Macro F1 \\
\hline
1 & M3 & 7 & 119 & 0.7000 & 0.1029 & 0.1040 \\
1 & M4 & 7 & 66  & 0.7000 & 0.1687 & 0.2501 \\
2 & M3 & 7 & 113 & 0.7000 & 0.1077 & 0.2153 \\
2 & M4 & 5 & 81  & 0.5000 & 0.1042 & 0.1523 \\
3 & M3 & 5 & 98  & 0.5000 & 0.0885 & 0.1217 \\
3 & M4 & 7 & 69  & 0.7000 & 0.1628 & 0.2316 \\
\hline
Mean & M3 & 6.33 & 110.0 & 0.6333 & 0.0997 & 0.1470 \\
Mean & M4 & 6.33 & 72.0  & 0.6333 & 0.1452 & 0.2113 \\
\hline
\end{tabular}
\end{table}

\subsection{Cross-Dataset External Mapping Validation}
\label{sec:cross_dataset_validation}

Across the three evaluated SWaT generations, DT enrichment reduces annotation-relative FP episodes, while recall varies between runs. To determine whether the behavior observed in the frozen primary realization transfers across telemetry regimes, the corresponding frozen M3 and M4 configurations are subsequently evaluated on BATADAL and WADI. No retrieval, mapping, confidence, or consolidation parameter is modified after observing the external datasets. The cross-dataset comparison uses the frozen primary SWaT outputs to remain consistent with the single frozen evaluation performed on each external dataset.

Table~\ref{tab:cross_dataset_mapping} compares the frozen M3 and M4 configurations across all three datasets. The result reveals that the SWaT advantage does not transfer uniformly. In the frozen primary SWaT run, M4 preserves the M3 recall of 0.70 while reducing annotation-relative FP episodes from 119 to 66, increasing micro F1 from 0.1029 to 0.1687. On BATADAL, M3 obtains four true positives and 120 FPs, corresponding to an F1 of 0.0615, while M4 obtains three true positives and 146 FPs, reducing F1 to 0.0387. On WADI, both variants recover the same three of eight primary ATT\&CK nodes and obtain the same recall of 0.375. M3 produces 215 FP episodes and an F1 of 0.0265, while M4 produces 226 FPs and an F1 of 0.0253.

\begin{table}[!htbp]
\centering
\caption{Cross-dataset episode-level ATT\&CK mapping F1 for the frozen primary M3 and M4 outputs. M4 improves over M3 without structural DT context on SWaT in this realization, but the improvement does not transfer to BATADAL or WADI.}
\label{tab:cross_dataset_mapping}
\small
\begin{tabular}{@{}llrrrrrr@{}}
\toprule
\textbf{Dataset} &
\textbf{Method} &
\textbf{TP} &
\textbf{FP} &
\textbf{FN} &
\textbf{Prec.} &
\textbf{Rec.} &
\textbf{F1} \\
\midrule
SWaT     & M3 & 7 & 119 & 3 & 0.0556 & 0.7000 & 0.1029 \\
SWaT     & M4 & 7 &  66 & 3 & 0.0959 & 0.7000 & 0.1687 \\
BATADAL  & M3 & 4 & 120 & 2 & 0.0323 & 0.6667 & 0.0615 \\
BATADAL  & M4 & 3 & 146 & 3 & 0.0201 & 0.5000 & 0.0387 \\
WADI     & M3 & 3 & 215 & 5 & 0.0138 & 0.3750 & 0.0265 \\
WADI     & M4 & 3 & 226 & 5 & 0.0131 & 0.3750 & 0.0253 \\
\bottomrule
\end{tabular}
\end{table}

The external results show that the improvement observed on SWaT should not be interpreted as an unconditional benefit of DT enrichment. Absolute precision remains low on both BATADAL and WADI, and M4 does not outperform M3 on either external dataset. Additional operational context can improve semantic attribution when it constrains already informative evidence, but it cannot compensate for missing adversary-specific observations and may instead broaden the set of plausible but unsupported ATT\&CK hypotheses.

The BATADAL result exposes failures at several stages of the pipeline. Of the six primary ground-truth nodes, five have an eligible DT evidence window and the correct ATT\&CK technique appears within the frozen top-8 candidate set for all five. M3 maps four of these nodes correctly, while M4 maps three. One documented reporting-manipulation target is lost before retrieval because the relevant DT evidence window is absent. The remaining misses occur after successful retrieval, showing that semantic evidence selection remains a source of error in this dataset. Sparse hourly telemetry also provides substantial process-level context without equivalent adversary-specific artifacts, which can allow DT-enriched reasoning to construct additional plausible but unsupported interpretations.

WADI exhibits a different failure pattern. Of the eight primary telemetry-evaluable ATT\&CK nodes, seven produce at least one eligible DT evidence window. The correct technique reaches the frozen top-8 candidate set for only three of these nodes. These are the three T1692.002 Reporting Message scenarios, and both M3 and M4 map all three correctly. None of the evaluable T1692.001 Command Message targets reaches the frozen top-8 candidate set, while T0836 Modify Parameter also remains outside the top eight. One command-message target, WADI Attack~12, is lost earlier because no eligible event window is generated for the corresponding control-output variable.

Consequently, the final WADI recall of 0.375 exactly matches the pre-inference candidate-availability ceiling. The end-to-end score alone hence understates where the failure occurs. The eight primary nodes are reduced to seven during DT evidence extraction and then to three during frozen ATT\&CK retrieval; all three targets surviving both stages are subsequently mapped correctly by both M3 and M4. Under the evaluated configuration, the dominant WADI limitation is upstream evidence and candidate availability rather than failure of the semantic reasoning stage to choose the correct technique once that technique is presented.

Table~\ref{tab:external_stage_coverage} summarizes this distinction. BATADAL retains five of six primary nodes through both DT evidence extraction and retrieval, after which M3 and M4 map four and three nodes, respectively. WADI follows a different pattern, with the largest loss occurring before semantic selection.

\begin{table}[!htbp]
\centering
\caption{Stage-wise coverage of primary ground-truth nodes in the external validation datasets. ``Top-8'' denotes appearance of the correct target in at least one eligible frozen retrieval candidate set.}
\label{tab:external_stage_coverage}
\small
\begin{tabular}{@{}lrrrrr@{}}
\toprule
\textbf{Dataset} &
\textbf{GT} &
\textbf{DT evidence} &
\textbf{Top-8} &
\textbf{M3 TP} &
\textbf{M4 TP} \\
\midrule
BATADAL & 6 & 5 & 5 & 4 & 3 \\
WADI    & 8 & 7 & 3 & 3 & 3 \\
\bottomrule
\end{tabular}
\end{table}

The external experiments qualify the SWaT result. DT structural context is beneficial when it constrains an already security-relevant interpretation, but additional operational information does not create adversary-specific evidence. In sparse or weakly observable settings, the same context can broaden the set of semantically plausible ATT\&CK hypotheses and increase FPs. The WADI experiment additionally demonstrates that downstream LLM reasoning cannot recover a documented behavior that is removed by an earlier DT-extraction or retrieval stage.

\subsection{Parameter Sensitivity and Robustness}
\label{sec:sensitivity}

Sensitivity analysis re-filters the frozen primary outputs without additional LLM inference. Across mapping thresholds $\theta_{\mathrm{map}}\in\{0.50,0.55,0.60,0.65\}$, M4 retains higher micro F1 than M3; at 0.70 both configurations reject all true-positive episodes. This loss indicates that the correctly matched predictions in the frozen outputs do not survive the higher confidence requirement. Increasing the threshold does not necessarily produce a useful high-precision operating point. The model-generated confidence values are not calibrated probabilities, and their ability to distinguish supported from unsupported mappings remains unestablished. Varying the episode-consolidation gap from 300 to 1800~s leaves recall at 0.70 for both methods in the frozen primary run and does not reverse the M4 advantage. Retrieval coverage reaches 10/10 SWaT targets at $k\geq8$, remains 5/6 for BATADAL across $k\in\{4,8,12,16\}$, and is 3/8 for WADI through $k=12$, increasing to 4/8 at $k=16$. The $k=16$ result measures candidate availability only; the LLM is not rerun with the enlarged set. These diagnostics are used to assess robustness, not to reselect the frozen configuration after observing held-out performance.

\subsection{Dependency Inference Results}
\label{sec:dependency_results}

\subsubsection{Documented Multi-Point Coordination}

The first dependency experiment isolates relation reasoning from upstream mapping errors by supplying oracle ATT\&CK nodes from five multi-point SWaT scenarios. The benchmark contains 12 documented attack-action nodes and nine positive Co-occurrence relations.

Table~\ref{tab:dependency_coordination} reports coordination recovery for D1--D4. D1 recovers all nine pairs because all documented actions satisfy the broad overlap heuristic. D4 recovers eight of the nine relations after receiving DT process and topology context, while D3 abstains on every pair when that structural context is withheld.

\begin{table}[!htbp]
\centering
\caption{Dependency inference on the frozen positive coordination benchmark. All nine ground-truth pairs are Co-occurrence relations.}
\label{tab:dependency_coordination}
\begin{tabular}{@{}lrrr@{}}
\toprule
\textbf{Method} &
\textbf{Correct} &
\textbf{Coordination Recall} &
\textbf{Abstention} \\
\midrule
D1 & 9/9 & 1.0000 & 0.0000 \\
D2 & 7/9 & 0.7778 & 0.2222 \\
D3 & 0/9 & 0.0000 & 1.0000 \\
\textbf{D4} & \textbf{8/9} & \textbf{0.8889} & \textbf{0.1111} \\
\bottomrule
\end{tabular}
\end{table}

The contrast between D3 and D4 demonstrates that DT structural context provides information that the LLM uses to establish operational relatedness. Without process-stage, control-role, and topology information, D3 does not infer any of the documented coordinated relationships. D4 recovers eight of nine once this context is supplied.

This result does not demonstrate superiority over simple temporal reasoning. D1 recovers all nine pairs because the documented actions within each scenario share broad attack intervals and satisfy the overlap heuristic. The positive benchmark is consequently useful as a coordination-recovery sanity check, but it is too temporally homogeneous to establish general relation-classification performance.

The single D4 miss occurs in A30 between P101 and MV101. The model recognizes that both actions overlap and are operationally connected, but returns None because the available topology suggests a possible direction opposite to the ordered input pair. This error illustrates a tension between representing Co-occurrence as a symmetric relation and requesting a relation ``from A to B'' in the dependency prompt. The frozen output is retained without retuning.

\subsubsection{End-to-End Edge Specificity}

The second dependency experiment evaluates relation inference in the presence of upstream mapping errors. The frozen M4 mapping output across the nine held-out scenarios contains 73 predicted episodes, of which seven are mapping true positives and 66 are FPs. Candidate dependency pairs are generated from these episodes under the frozen 900\,s temporal gap. The final exploratory benchmark contains one telemetry-evaluable positive dependency and 25 annotation-relative negative-control pairs.

Table~\ref{tab:e2e_dependency} reports edge-existence performance on the selected 26-pair benchmark. D1 connects every pair, recovering the sole positive relationship while rejecting none of the 25 negative controls. D3 rejects 21 negative controls but misses the positive relationship. D4 recovers that relationship and rejects 12 negative controls, producing 13 false edges and specificity 0.48. Positive-edge recall is determined by one observation, making recall and balanced accuracy unsuitable as stable estimates of general performance. Specificity characterizes the 25 selected negative controls and should not be interpreted as a deployment-level estimate.

\begin{table*}[!htbp]
\centering
\caption{Exploratory end-to-end dependency evaluation. The benchmark contains one positive edge and 25 annotation-relative negative-control pairs. Specificity and FPR are the primary metrics.}
\label{tab:e2e_dependency}
\resizebox{\linewidth}{!}{%
\begin{tabular}{@{}lrrrrrrrrr@{}}
\toprule
\textbf{Method} &
\textbf{TP} &
\textbf{FP} &
\textbf{TN} &
\textbf{FN} &
\textbf{Edge Prec.} &
\textbf{Edge Rec.} &
\textbf{Specificity} &
\textbf{FPR} &
\textbf{Bal. Acc.} \\
\midrule
D1 & 1 & 25 & 0  & 0 & 0.0385 & 1.0000 & 0.0000 & 1.0000 & 0.5000 \\
D2 & 0 & 13 & 12 & 1 & 0.0000 & 0.0000 & 0.4800 & 0.5200 & 0.2400 \\
D3 & 0 & 4  & 21 & 1 & 0.0000 & 0.0000 & \textbf{0.8400} & \textbf{0.1600} & 0.4200 \\
\textbf{D4} & \textbf{1} & 13 & 12 & \textbf{0} & 0.0714 & \textbf{1.0000} & 0.4800 & 0.5200 & \textbf{0.7400} \\
\bottomrule
\end{tabular}
}
\end{table*}

Because only one observable positive edge is available, D4's edge recall and balanced-accuracy values should not be interpreted as stable estimates of general positive-edge performance. The end-to-end experiment instead reveals a trade-off that is not visible in the positive-only coordination benchmark. DT context substantially increases the LLM's willingness to infer operational relationships, but the additional sensitivity also allows erroneous upstream ATT\&CK hypotheses to become connected by plausible-looking graph edges.

\subsection{Dependency Error Analysis}
\label{sec:dependency_error_analysis}

The error analysis groups candidate pairs by interval overlap and operational relatedness. Overlap-operational pairs have overlapping episode intervals and an operational relationship identified by the DT; sequential-operational pairs have non-overlapping intervals and an operational relationship. Overlap-only pairs overlap without such a relationship, whereas sequential-unrelated pairs are non-overlapping and lack an identified operational relationship. These categories describe the supplied context and do not establish adversarial dependency. 

The 13 false edges produced by D4 are not distributed randomly. Six are labeled Temporal, four Co-occurrence, and three Causal. More importantly, 12 of the 13 false edges occur between pairs for which the DT exposes explicit operational coherence. Eight occur in the \emph{sequential-operational} group and four in the \emph{overlap-operational} group. Only one false edge occurs for an \emph{overlap-only} pair, and no false edge is produced for a \emph{sequential-unrelated} pair.

Inspection of the generated rationales shows a consistent failure pattern. For false Temporal edges, the model combines temporal proximity with shared assets, same-stage membership, or an upstream/downstream DT relation and interprets this combination as attack progression. For false Co-occurrence edges, simultaneous activity on operationally related assets is treated as evidence that the corresponding ATT\&CK hypotheses belong to the same malicious action sequence. The causal errors are more consequential: in these cases, the model constructs a plausible process-propagation argument and then interprets that physical or operational pathway as evidence that one attack action caused or enabled another.

The false-edge composition also demonstrates propagation of mapping errors into the graph. Nine of the 13 D4 false edges connect two mapping FP nodes, while the remaining four connect one true-positive mapping node and one FP node. None connects two correctly mapped nodes. Thus, the dominant failure mechanism is: mapping error $\rightarrow$ plausible DT context $\rightarrow$ unsupported graph edge.

These observations establish an important limitation of DT-guided semantic reasoning. Operational connectivity is useful for deciding which events deserve joint analysis, but operational connectivity is not itself evidence of adversarial dependency. In particular, temporal proximity plus operational coherence does not imply attack progression.

Likewise, a process propagation path does not establish attacker-induced causality. Confirming Causal edges requires stronger evidence, such as command provenance, explicit attacker actions, state transitions attributable to those actions, or corroborating observations from additional security telemetry.

\subsection{Case Study: SWaT A39 Graph Reconstruction}
\label{sec:a39_case_study}

SWAT\_A39 provides the clearest end-to-end example because both telemetry-evaluable attack actions are correctly mapped. The frozen mapping output contains four predicted ATT\&CK episodes. Two are true positives corresponding to T1692.002 (\emph{Reporting Message}): one associated with FIT401 in process stage P4 and one associated with AIT502 in P5. The remaining two predicted episodes are mapping FPs.

The documented A39 actions involving FIT401 and AIT502 form a coordinated multi-point manipulation and are annotated with a Co-occurrence dependency. In the frozen end-to-end dependency experiment, D4 correctly determines that an edge should exist between the two true-positive nodes, but labels the relation as Causal with confidence 0.72. The edge-existence decision is operationally plausible while the relation semantics are overstated. The DT provides a process path connecting the corresponding stages and measurements, but this process connectivity does not demonstrate that manipulation of FIT401 caused manipulation of AIT502.

\begin{figure}[pos=htbp]
\centering
\includegraphics[width=\linewidth]{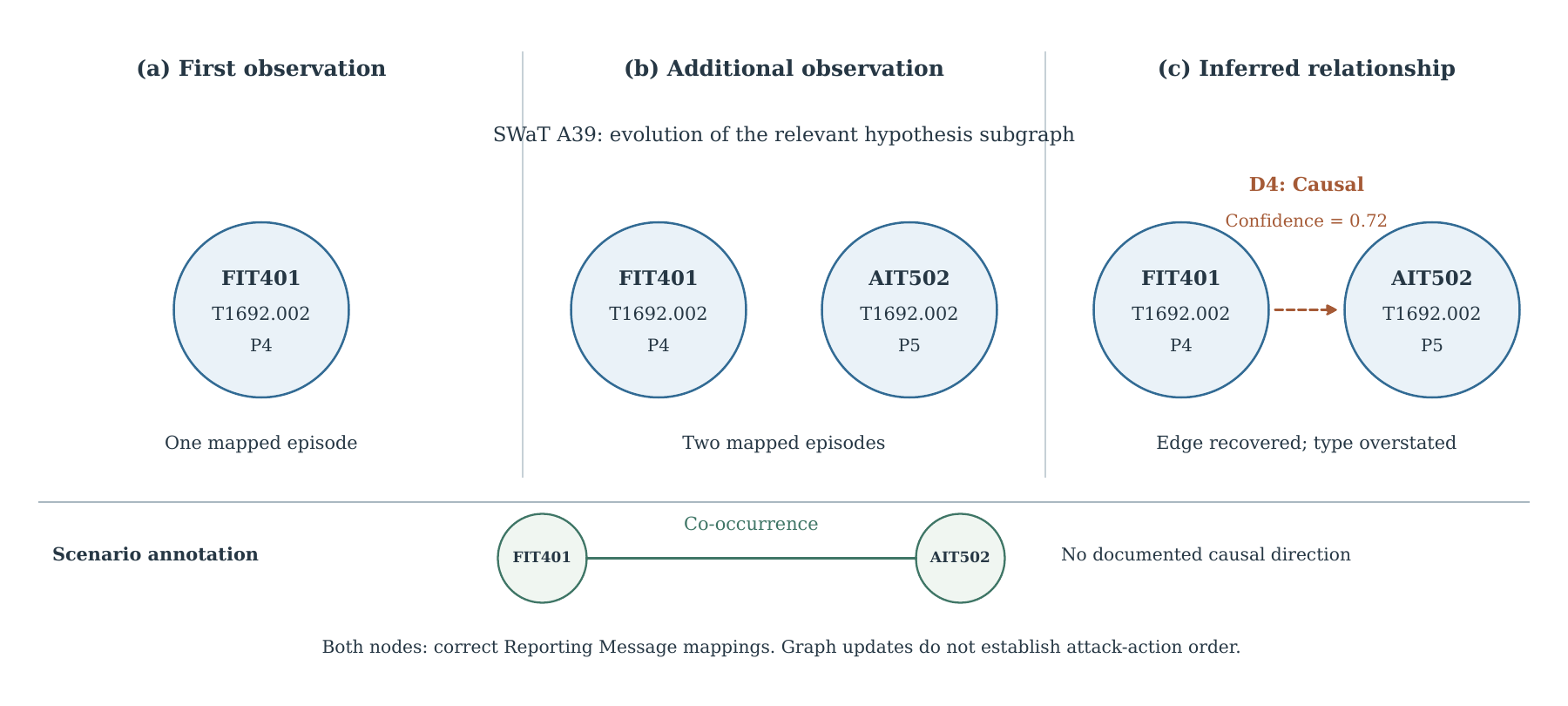}
\caption{Evolution of the relevant SWaT A39 hypothesis subgraph.}
\label{fig:casestudy}
\end{figure}

Figure~\ref{fig:casestudy} shows successive graph updates for the two correctly mapped episodes; the two FP episodes are omitted. The first state contains the FIT401 hypothesis, the second adds AIT502, and the final state introduces the inferred relationship. These updates reflect evidence availability and not a verified ordering of attacker actions. D4 recovers the relationship but predicts Causal with confidence 0.72, whereas the scenario documentation supports only Co-occurrence.

% ═══════════════════════════════════════════════════════════════════════════════
\section{Discussion}
\label{sec:dis}

The principal SWaT observation is lower annotation-relative FP counts under DT enrichment in each of the three evaluated generations. However, recall varies across runs, the primary-run paired comparisons do not reach statistical significance, and M4 does not improve F1 in every generation. The result provides limited descriptive evidence of increased selectivity under the evaluated conditions. Its external applicability is further constrained by the BATADAL and WADI results, where M4 does not outperform M3. These findings support investigation of when DT context constrains semantic interpretation, without establishing a general mapping advantage.

Dependency reasoning exposes the complementary risk. DT topology and process semantics make operationally coherent observations easier to relate, but the same context can connect incorrect mapping hypotheses through plausible Temporal, Co-occurrence, or Causal edges. The A39 case illustrates this boundary directly: the existence of a relationship is plausible, while the predicted Causal semantics exceed the documented Co-occurrence evidence. Therefore, DT context should prioritize and contextualize hypotheses, not serve as independent proof of attacker-induced causality. Stronger directional claims require corroboration from command provenance, network-message evidence, authentication/security logs, or multiple mutually consistent telemetry sources. Future systems should propagate extraction, retrieval, observability, mapping, and edge uncertainty explicitly across stages.

The mapping ablation compares LLM-based configurations and does not include a non-generative retrieval baseline. It does not establish whether LLM reasoning improves on direct selection from the retrieved candidates. Practical analyst benefit also remains unevaluated. In the frozen primary outputs, M4 produces 66 FP episodes for seven true positives on SWaT, 146 for three on BATADAL, and 226 for three on WADI, corresponding to approximately 9.4, 48.7, and 75.3 annotation-relative false episodes per true episode. These ratios indicate substantial potential review burden, although they do not measure analyst workload because episode redundancy, review time, and evidence quality are not evaluated. An analyst-facing implementation would require assessment of prioritization, abstention, and investigation cost before operational usefulness could be established.

Several threats to validity remain. ATT\&CK and dependency ground truth are reconstructed conservatively from scenario documentation and available telemetry and are annotation-relative and not exhaustive. Historian data expose process state more reliably than adversary provenance, and the implemented twin is a state-based cybersecurity DT instead of a high-fidelity hydraulic causal model. Three M3/M4 runs expose generation variability without estimating the full output distribution, while dependency classification itself is not repeated. All evaluated datasets are water-sector CPSs, so external validation demonstrates transfer across telemetry regimes and not industrial domains in general. The positive dependency benchmark contains only Co-occurrence relations, while the end-to-end benchmark contains one observable positive edge and 25 selected annotation-relative negative controls. Therefore, its specificity characterizes the selected pairs rather than all 381 candidate pairs or an operational deployment. These experiments support analysis of coordination recognition and error propagation, but do not establish general Temporal or Causal relation-classification performance.

\section{Conclusion}
\label{sec:con}

This paper presented a DT-assisted framework for mapping ICS telemetry to ATT\&CK for ICS observations and organizing those observations within a typed dependency graph. Across three evaluated SWaT generations, DT enrichment reduced annotation-relative FP episodes in every run, but recall varied and F1 did not improve consistently. Primary-run paired comparisons did not reach statistical significance, and the mapping advantage did not transfer to BATADAL or WADI. The exploratory dependency experiments showed that DT context can support recognition of coordinated observations while also connecting incorrect mappings through plausible operational relationships. The contribution is therefore an empirical characterization of the opportunities and failure modes of DT-assisted semantic reasoning. Further evaluation should isolate the contribution of generative reasoning, strengthen annotation validation, expand generation and dependency coverage, and measure the practical cost of reviewing the resulting hypotheses.

\section*{CRediT authorship contribution statement}
\textbf{Konstantinos E. Kampourakis:} Conceptualization, Methodology, Software, Validation, Writing – original draft, Writing – review \& editing.
\textbf{Vyron Kampourakis:} Writing – original draft, Writing – review \& editing.
\textbf{Vasileios Gkioulos:} Writing – review \& editing, Supervision.
\textbf{Sokratis Katsikas:} Writing – review \& editing, Supervision.

\section*{Funding}
This work is supported by the Research Council of Norway through
the SFI Norwegian Centre for Cybersecurity in Critical Sectors (NORCICS) project no. 310105.

\section*{Declaration of competing interest}
The authors declare that they have no known competing interests.

\section*{Data and code availability}
Data will be made available on request.

\bibliographystyle{elsarticle-num}
\bibliography{references}

\end{document}